\documentclass[12pt]{article}%\documentstyle[12pt]{article}
\usepackage{amsmath}
\usepackage{graphicx}%
\usepackage{amsfonts}%
\usepackage{amssymb}
\usepackage{psfrag}

\newcommand{\bea}{\begin{eqnarray}}
\newcommand{\eea}{\end{eqnarray}}
\newcommand{\be}{\begin{equation}}
\newcommand{\ee}{\end{equation}}
\newcommand{\vs}[1]{\vspace{#1 mm}}

\newcommand{\dsl}{\pa \kern-0.5em /}

\newcommand{\pa}{\partial}

\newcommand{\nn}{\nonumber\\}

\newcommand{\eqn}[1]{(\ref{#1})}

\begin{document}
\topmargin 0pt
\oddsidemargin 0mm

\begin{flushright}

USTC-ICTS-09-07\\

\end{flushright}

\vspace{2mm}

\begin{center}

{\Large \bf Remarks on D$_p$ $\&$ D$_{p - 2}$ with each carrying a
flux } \vs{6}

{\large J.X. Lu\footnote{E-mail: jxlu@ustc.edu.cn} and Shan-Shan
Xu\footnote{E-mail: xuss@mail.ustc.edu.cn} }

\vspace{4mm}

{\em

 Interdisciplinary Center for Theoretical Study\\

 University of Science and Technology of China, Hefei, Anhui
 230026, China\\}

\vs{4}

\end{center}

%\vs{10}

\begin{abstract}
We explore various properties of interaction between D$_p$ and
D$_{p'}$ with each carrying a worldvolume flux and with the
$p'$-branes placed along  $p'$ spatial worldvolume  directions of
the p-branes at a separation when $p - p' = 2$. Carefully analyzing
the annulus amplitudes calculated via the boundary state approach,
we find that many features of amplitudes remain similar to those
studied when $p = p'$ such as the nature of force on the brane
separation, the onset of various instabilities when the
brane-separation is on the order of string scale and the occurrence
of pair production of open strings when there is a relevant electric
flux present. In addition, we have also found many new features of
interaction which don't appear in the absence of fluxes or when $p =
p'$ in the presence of fluxes, for examples, the nature of
interaction can be repulsive and there is no onset of tachyonic
instability under certain conditions. Even in the absence of a
magnetic flux, we can have an exponential enhancement of the rate of
pair production of open strings in certain cases, which may be
significant enough to give observational consequences.

\end{abstract}
\newpage

\noindent It is known that the static interaction between a D$_p$
and a D$_{p'}$ brane at a separation with $p - p' = 2$ is always
attractive\footnote{In other words, we consider in this paper the
case of $p - p' = 2$ with $NN = p' + 1, ND = 2$ and $DD = 9 - p$.
The D$_p$ and D$_{p'}$ are placed at a separation along the $DD = 9
- p$ directions with $p \le 8$.} when neither carries a world-volume
flux, for example, see \cite{Di Vecchia:1999rh}. This nature of
interaction is expected to be modified or changed when there is an
electric or a magnetic flux present on either of the
branes\footnote{As mentioned in \cite{polbooktwo, Lu:2009kv2} and
the references therein, D$_k$ branes become a
1/2$-$Bogomol'nyi-Prasad-Sommerfield (BPS) non-threshold (F, D$_k$)
bound state when carrying an electric flux while a 1/2-BPS
non-threshold (D$_{k - 2}$, D$_k$) bound state when carrying a
magnetic flux.}. In this note, we will explore this modification or
change and the associated properties such as the onset of various
instabilities and the open string pair production, expecting some
unique and interesting physical implications to arise. We find that
the pair production rate is significant even for the mere presence
of a weak electric flux under certain condition. Unlike the rate
enhancement discussed in \cite{Lu:2009kv2}, an additional magnetic
flux not sharing any common index with the electric flux is not
necessarily needed. The novel feature here is that the D$_{p -2}$
brane itself takes the role of the magnetic flux to enhance the
rate. The magnetic flux can further either enhance or reduce the
rate, depending on the actual case considered. One interesting
observation is that the largest rate occurs either for $p = 3$ as
also observed in \cite{Lu:2009kv2}, for which $p = p'$ was
considered, or for $p' = 3$. We don't know if this has an
implication of why our world has three large spatial dimensions and
even for the existence of extra dimensions. When both fluxes are
magnetic, the would-be attractive interaction can vanish in certain
cases under conditions specified later, signalling the preservation
of certain number of supersymmetries (susy) of the underlying
systems. When both these fluxes point to different NN directions,
these is an interesting case for which the force can even be
repulsive for certain range of the fluxes. For this case and the
above susy cases, the corresponding amplitude doesn't give rise to a
tachyonic instability, therefore no tachyon condensation to occur.

Without further ado, we have the following three cases to
consider\footnote{In this note, the Greek indices $\alpha = (0, a),
\beta = (0, b), \cdots$ label the world-volume directions $0, 1,
\cdots, p$ along which the D$_p$ brane extends with $a, b, \cdots$
denoting the brane spatial diretions, while the later Latin indices
$i, j, \cdots$ label the directions transverse to the brane, i.e.,
$p + 1, \cdots, 9$.}: 1) D$_p$ and D$_{p'}$ carry their respective
electric fluxes $F_{0a}$ and $F'_{0b}$; 2) D$_p$ and D$_{p'}$ carry
their respective magnetic fluxes $F_{ab}$ and $F'_{cd}$; 3) one
carries an electric flux and the other carries a magnetic flux. Or
we can classify the interaction amplitudes according to their
structures determined by the relative orientations of the two fluxes
$F_{\alpha\beta}$ and $F'_{\gamma\delta}$ in the following three
classes: I) the indices $\alpha, \beta, \gamma, \delta \in {\rm NN}$
with the pair ($\alpha,\, \beta$) and the pair ($\gamma, \, \delta$)
sharing at least one common index (either temporal or spatial) or
the index $\alpha$ or $\beta \in {\rm ND}$ but not both; II)
$\alpha, \beta, \gamma, \delta \in {\rm NN}$ but $\alpha, \beta \neq
\gamma, \delta$; III) $\alpha, \beta \in {\rm ND}$ and $\gamma,
\delta \in {\rm NN}$. Note that we have $p \ge 3$  for Class I or
III and  $p \ge 5$ for Class II. We will present the amplitudes
according to this classification for simplicity and for a unified
description in each class. Note that we have  non-vanishing
contributions only from the Neveu-Schwarz$-$Neveu-Schwarz (NS-NS)
sector for amplitudes in Class I or Class II but this is not true
for amplitudes in Class III.

 The  tree-level cylinder
diagram interaction can be calculated via the boundary state
approach. As usual, we have two sectors, namely NS-NS and
Ramond$-$Ramond (R-R) sectors, respectively. We here summarize the
main results needed, following \cite{DiVecchia:1999uf}. Let $F$ be
the external flux on the world-volume and denote $\hat{F}=2\pi
\alpha^{\prime}F$. In the NS-NS sector, the relevant boundary state
is the Gliozzi-Scherk-Olive (GSO) projected one
$|B\rangle_{NS}=\frac{1}{2}\left[|B,+\rangle_{NS}-|B,-\rangle_{NS}\right]$
while in the R-R sector, the GSO projected one is
$|B\rangle_{R}=\frac{1}{2}\left[|B,+\rangle_{R}+|B,-\rangle_{R}\right]$.
Here the two boundary states $|B, \eta\rangle$ with $\eta = \pm$
correspond to two possible implementations for the boundary
conditions in each sector.  The boundary state $|B, \eta\rangle$ is
the product of a matter part and a ghost part as $|B, \eta\rangle =
\frac{c_p}{2} |B_{\rm mat.}, \eta\rangle |B_g, \eta\rangle$
 with $|B_{\rm mat.}, \eta\rangle = |B_X\rangle
|B_\psi, \eta\rangle$, $|B_g,\eta\rangle = |B_{gh}\rangle |B_{\rm
sgh}, \eta\rangle$ and the normalization constant
$c_p=\sqrt{\pi}\left(2\pi\sqrt{\alpha^{\prime}}\right)^{3-p}$. The
explicit forms of the various components of $|B\rangle$ are given as
$ |B_X\rangle = {\rm exp}[-\sum_{n =1}^\infty \frac{1}{n} \alpha_{-
n} \cdot S \cdot {\tilde \alpha}_{ - n}] |B_X\rangle^{(0)}$, and for
the NS-NS sector $|B_\psi, \eta\rangle_{\rm NS} = -  {\rm i} ~ {\rm
exp} [ {\rm i}\, \eta \sum_{m = 1/2}^\infty \psi_{- m} \cdot S \cdot
{\tilde \psi}_{- m} ] |0\rangle $  and for the R-R sector $ |B_\psi,
\eta\rangle_{\rm R} = - {\rm exp} [{\rm i}\, \eta \sum_{m =
1}^\infty \psi_{- m} \cdot S \cdot {\tilde \psi}_{- m} ] |B,
\eta\rangle_{\rm R}^{(0)} $. The matrix $S$ and the zero-mode
contributions $|B_X\rangle^{(0)}$ and $|B, \eta\rangle_{\rm
R}^{(0)}$ encode all information about the external flux and the
overlap equations that the string coordinates have to satisfy, which
in turn depend on the boundary conditions of the open strings ending
on the D-brane. They each can be given explicitly as $ S = ([(\eta -
\hat{F})(\eta + \hat{F})^{-1}]_{\alpha\beta}, - \delta_{ij})$, $
|B_X\rangle^{(0)} = \sqrt{- \det (\eta + \hat F)} \,\delta^{9 - p}
(q^i - y^i) \prod_{\mu = 0}^9 |k^\mu = 0\rangle$, and $ |B_\psi,
\eta\rangle_{\rm R}^{(0)} = (C \Gamma^0 \Gamma^1 \cdots \Gamma^p
\frac{1 + {\rm i}\, \eta \Gamma_{11}}{1 + {\rm i}\, \eta } U )_{AB}
|A\rangle |\tilde B\rangle$.   In the above, we have denoted by
$y^i$ the positions of the D-brane along the transverse directions,
by $C$ the charge conjugation matrix and by $U$ the following matrix
$\label{umatrix} U =  \,; {\rm exp}(- \frac{1}{2} {\hat
F}_{\alpha\beta}\Gamma^\alpha\Gamma^\beta);\,/\sqrt{- \det (\eta +
\hat F)}$ where the symbol $;\quad ;$ means that one has to expand
the exponential and then to anti-symmetrize the indices of the
$\Gamma$-matrices. $|A\rangle |\tilde B\rangle$ stands for the
spinor vacuum of the R-R sector. Note that the $\eta$ in the above
means either sign $\pm$ or the flat signature matrix $(-1, +1,
\cdots, +1)$ on the world-volume and should not be confused from the
content. Note also that the boundary state must be written in the
$(-1, -1)$ super-ghost picture in the NS-NS sector, and in the
asymmetric $(- 1/2, - 3/2)$ picture in the R-R sector in order to
saturate the super-ghost number anomaly of the disk.

The interaction under consideration can be calculated as the vacuum
amplitude of the closed string tree-level cylinder diagram via the
boundary states as just described. The total amplitude can have
contributions from both NS-NS sector and R-R sector and is given by
$\Gamma = \Gamma_{\rm NS} + \Gamma_{\rm R}$ with $ \label{4eq1}
\Gamma_{\rm NS/R} =\,  _{\rm NS/R}\langle
B_{p-2}|D|B_{p})\rangle_{\rm NS/R}$. Here $D$ is the closed string
propagator
\begin{equation}
\label{4eq2}
 D=\frac{\alpha^{\prime}}{4\pi}\int_{|z|\leq1}\frac{d^2z}{|z|^2}z^{L_0}\bar{z}^{\tilde{L}_0},
\end{equation}
with $L_0$ and $\tilde{L}_0$ the respective left and right moving
total zero-mode Virasoro generators of matter fields, ghosts and
superghosts. For example, $L_0 = L^X_0+L_0^\psi+L_0^{gh}+ L_0^{sgh}$
and their explicit expressions can be found, e.g.,  from
\cite{polbooktwo}. The above amounts to calculating first the
following amplitude in the respective sector
\begin{eqnarray}
\label{5eq1} \Gamma(\eta',\eta) = \langle
B_{p-2},\eta'|D|B_{p},\eta\rangle = \frac{n_{p - 2} n_p
c_{p-2}c_p}{4} \frac{\alpha'}{4 \pi} \int_{|z| \le 1} \frac{d^2
z}{|z|^2} A^X \, A^{bc}\, A^\psi (\eta', \eta)\, A^{\beta\gamma}
(\eta', \eta)
\end{eqnarray}
 where $\eta'\eta = \pm$ and we have also
replaced the $c_k$ ($k = p - 2 \, {\rm or}\, p$) in the boundary
state given earlier by $n_k c_k$ with $n_k$ an integer to count the
multiplicity of the D$_k$ branes in the bound state. In the above,
the various matrix elements are
\begin{eqnarray}
\label{5eq2} && A^X = \langle B^X_{p-2} | |z|^{2L_0^{X}}|B^X_p
\rangle,\,\,\,\, A^\psi (\eta', \eta) = \langle B^\psi_{p-2}, \eta'|
|z|^{2L_0^{\psi}} |B^\psi_{p}, \eta \rangle,\nn && A^{bc} = \langle
B^{gh}_{p-2} | |z|^{2 L_0^{bc}} |B^{gh}_{p} \rangle, \,\,\,\,
A^{\beta\gamma} (\eta', \eta) = \langle B^{sgh}_{p-2}, \eta'| |z|^{2
L_0^{\beta\gamma}} |B^{sgh}_{p}, \eta \rangle.
\end{eqnarray}
where we have used the boundary state constraint $\tilde L_0
|B\rangle = L_0 |B\rangle$ to simplify the calculations. Note that
the $A^{bc}$ and $A^{\beta\gamma}$ are independent of fluxes and are
always given as $\label{4eq3} A^{bc}=
|z|^{-2}\prod_{n=1}^{\infty}\left(1-|z|^{2n}\right)^2$ in both NS-NS
and R-R sectors, and in the NS-NS sector, $\label{4eq4} A_{\rm
NS}^{\beta\gamma}(\eta',\eta)=
|z|\prod_{n=1}^{\infty}\left(1+\eta'\eta|z|^{2n-1}\right)^{-2}$
while in the R-R sector, $\label{4eq5} A_{\rm
R}^{\beta\gamma}(\eta',\eta)=
|z|^{3/4}(1+\eta'\eta)^{-1}\prod_{n=1}^{\infty}\left(1-\eta'\eta|z|^{2n}\right)^{-2}$.

However in what follows, we will adopt the prescription given in
\cite{Billo:1998vr,Di Vecchia:1999rh} not to separate the
contributions from  matter fields $\psi^{\mu}$ and superghosts in
the R-R sector  to avoid the complication due to the respective zero
modes if this sector has non-zero contribution. To calculate $A^X$
and $A^\psi(\eta',\eta)$, we will follow the trick as described in
\cite{Lu:2009yx,Lu:2009kv2} by making a respective unitary
transformation of the oscillators in $|B^{X/\psi}_{p -
2},\eta'\rangle$ such that the $S_{p - 2}$-matrix there completely
disappears while $|B^{X/\psi}_p, \eta\rangle$ ends up with a new $S
= S_p S^T_{p - 2}$ with $S_p$ the original S-matrix in this boundary
state and $T$ denoting the transpose. This new S-matrix shares the
same property as the original $S_k$ satisfying $(S_k^T)_\mu\,^\rho
(S_k)_\rho\,^\nu = \delta_\mu\,^\nu$ with $k = p - 2\, {\rm or}\, p$
but its determinant is always unity and therefore can always be
diagonalized to gives its eigenvalues. With this trick, the
evaluation of $A^{X/\psi}$ is no more complicated than the case
without the presence of fluxes.

If we take $(\hat F')_{\gamma\delta} = - (\hat F')_{\delta\gamma} =
- f_1$ with $\gamma < \delta$ and $(\hat F)_{\alpha\beta} = - (\hat
F)_{\beta\alpha} = - f_2$ with $\alpha < \beta$, the Class I matrix
elements for matter fields are
\begin{equation}
\label{5eq7} A^X = C_F V_{p-1} e^{- \frac{Y^2}{2\pi \alpha' t}}
\left(2\pi^2 \alpha'\, t\right)^{- \frac{9 - p}{2}} \, \prod_{n =
1}^\infty \frac{1}{(1 - \lambda |z|^{2n})(1 - \lambda^{-1}
|z|^{2n})(1 + |z|^{2n})^2(1 - |z|^{2n})^6}
\end{equation}
for both NS-NS and R-R sectors,
\begin{equation}
\label{5eq8} A^\psi_{NS}(\eta',\eta) =\prod_{n = 1}^\infty (1 +
\eta' \eta\lambda |z|^{2n - 1}) (1 + \eta' \eta\lambda^{-1}\,
|z|^{2n - 1})(1 - \eta' \eta |z|^{2n - 1})^2 (1 + \eta' \eta |z|^{2n
- 1})^6
\end{equation}
for the NS-NS sector, and $A^\psi_{R}(\eta',\eta)=0$ for the R-R
sector. Note that $\Gamma_{\rm R} = 0$, so the total amplitude is
just
\begin{eqnarray} \label{7eq1}
\Gamma_{\rm I} = \Gamma_{\rm NS} &=&\frac{2 n_{p-2}\, n_p
\,V_{p-1}\, C_F
\,\sin\pi\nu}{(8\pi^2\alpha^{\prime})^{\frac{p-1}{2}}}
\int_{0}^{\infty}dt\,
e^{-\frac{Y^2}{2\pi\alpha^{\prime}t}}\,t^{-\frac{9-p}{2}}\,\frac{\theta^2_1(\frac{2\nu-1}{4}|it)\,\theta^2_1(\frac{2\nu+1}{4}|it)}
{\eta^6(it)\,\theta_1(\nu|it)\, \theta_1(\frac{1}{2}|it)},
\end{eqnarray}
where we have taken  $|z|=e^{-\pi t}$ and $\lambda = e^{2 i \pi
\nu}$.  In obtaining the above compact expression, we have first
expressed the integrand in terms of various $\theta$-functions and
the Dedekind $\eta$-function,  then made use of the fundamental
Jacobian identity $2\, \theta_1^2 (\frac{2\nu - 1}{4} | it) \,
\theta_1^2 (\frac{2 \nu + 1}{4} | it) = \theta_3 (\nu | it)\,
\theta_3 (\frac{1}{2} | it)\, \theta^2_3 (0 | it) - \theta_4 (\nu |
it) \, \theta_4 (\frac{1}{2} | it)\, \theta_4^2 (0 | it)$, which is
a special form of (iv) given on page 468 in \cite{whittaker-watson}.
The constant $C_F$ and the sum\footnote{We need only this sum to
determine $\nu$ via $\cos\pi\nu$ in terms of flux/fluxes given in
the table. When the flux/fluxes are electric or electrically
dominant, the $\nu$ is imaginary and  $\cos\pi\nu$ in the table
represents actually $\cosh\pi\nu_0$ since $\cos\pi\nu =
\cosh\pi\nu_0$ when $\nu = i \nu_0$. This remains also true in Table
2 and 3.} of $\lambda + \lambda^{ -1}$ can be summarized for cases
considered in this class in Table 1.

The Class II matrix elements for matter fields are
\begin{equation}
\label{5eq9} A^X = C_F V_{p-1} e^{- \frac{Y^2}{2\pi \alpha' t}}
\left(2\pi^2 \alpha'\, t\right)^{- \frac{9 - p}{2}} \, \prod_{n =
1}^\infty \frac{1}{(1 + |z|^{2n})^2(1 - |z|^{2n})^4} \prod_{j = 1}^2
\frac{1}{(1 - \lambda_j |z|^{2n})(1 - \lambda_j^{-1} |z|^{2n})}
\end{equation}
for both NS-NS and R-R sectors,
\begin{equation}
\label{5eq10} A^\psi_{NS}(\eta',\eta) =\prod_{n = 1}^\infty (1 -
\eta' \eta |z|^{2n - 1})^2 (1 + \eta' \eta |z|^{2n - 1})^4 \prod_{j
= 1}^2(1 + \eta' \eta\lambda_j |z|^{2n - 1}) (1 + \eta'
\eta\lambda_j^{-1}\, |z|^{2n - 1})
\end{equation}
for the NS-NS sector, and again $A^\psi_{R}(\eta',\eta)=0$ for the
R-R sector. Once again $\Gamma_{\rm R} = 0$ and the total amplitude
in this Class is
\begin{eqnarray} \label{7eq2}
\Gamma_{\rm II} = \Gamma_{\rm NS}
&=&\frac{4n_{p-2}n_pV_{p-1}\tan\pi\nu_1\tan\pi\nu_2}{(8\pi^2\alpha^{\prime})^{\frac{p-1}{2}}}
\int_{0}^{\infty}dt
e^{-\frac{y^2}{2\pi\alpha^{\prime}t}}t^{-\frac{9-p}{2}}\nn &\,&
\times
\frac{\theta_1(\frac{\nu_1-\nu_2-1/2}{2}|it)\theta_1(\frac{\nu_1-\nu_2+1/2}{2}|it)
\theta_1(\frac{\nu_1+\nu_2-1/2}{2}|it)\theta_1(\frac{\nu_1+\nu_2+1/2}{2}|it)}
{\eta^3(it)\theta_1(\nu_1|it)\theta_1(\nu_2|it)\theta_1(\frac{1}{2}|it)},
\end{eqnarray}
where we have taken $|z| = e^{- \pi t}$ and $\lambda_j = e^{2 i \pi
\nu_j}$ with $j = 1, 2$. Similarly, we first express the integrand
via various $\theta$-functions and the Dedekind $\eta$-function,
then use another fundamental Jacobian identity $2 \,\theta_1
(\frac{\nu_1 - \nu_2 - 1/2}{2} | it)\theta_1 (\frac{\nu_1 - \nu_2 +
1/2}{2} | it)\, \theta_1 (\frac{\nu_1 + \nu_2 - 1/2}{2} | it) \,
\theta_1 (\frac{\nu_1 + \nu_2 + 1/2}{2} | it) = \theta_3 (\nu_1 |
it)\, \theta_3(\nu_2 | it)\, \theta_3 (1/2 | it)\, \theta_3 (0 | it)
- \theta_4 (\nu_1 |it)\, \theta_4 (\nu_2 | it) \, \theta_4 (1/2 |
it) \,  \theta_4 (0 | it)$, which is also a special form of (iv)
given on page 468 in \cite{whittaker-watson}. The constant $C_F$ and
the sum of $\lambda_j + \lambda^{-1}_j$  for the cases considered in
this class are listed in Table 2.
\begin{table}
\caption{The cases in Class I} \vskip 0.2cm
\begin{tabular}{|c|c|c|c|c|c|}
\hline

$(\alpha, \beta)$& $(\gamma, \delta)$ & Index Relation  & $C_F$ &
$\frac{\lambda + \lambda^{-1}}{2}$ &$\cos\pi\nu$ \\
\hline

$(0, a)$ & $(0, c)$ & $a = c \in {\rm NN}$ & $\sqrt{(1 - f_1^2)(1 -
f_2^2)}$ & $\frac{(1 + f_1^2) (1 + f_2^2) - 4 f_1 f_2}{(1 - f_1^2)(1
- f_2^2)}$ & $\frac{1 - f_1 f_2}{C_F}$\\
\hline

$(0, a)$ & $(0, c)$ & $a, c \in {\rm NN}, \, a \neq c$ & $\sqrt{(1 -
f_1^2)(1 - f_2^2)}$ & $\frac{1 + f_1^2 + f_2^2 - f_1^2 f_2^2}{(1 -
f_1^2)(1
- f_2^2)}$& $\frac{1}{C_F}$ \\
\hline

$(0, a)$ & $(0, c)$ & $a \in {\rm DN}, \, c \in {\rm NN}$ &
$\sqrt{(1 - f_1^2)(1 - f_2^2)}$ & $\frac{1 + f_1^2 - f_2^2 + f_1^2
f_2^2}{(1 - f_1^2)(1
- f_2^2)}$ & $\frac{[1 - f_2^2(1 - f_1^2)]^{\frac{1}{2}}}{C_F}$\\
\hline

 $(0, a)$ & $(c, d)$ & $a, c, d \in {\rm NN},$   & $\sqrt{(1 + f_1^2)(1 - f_2^2)}$ & $\frac{1 - f_1^2
+ f_2^2 + f_1^2 f_2^2}{(1 + f_1^2)(1
- f_2^2)}$&$\frac{1}{C_F}$\\
 & & $a = c \,
{\rm or}\, d $& & &\\ \hline

$(0, a)$ & $(c, d)$ & $a \in {\rm DN}, \,c, d \in {\rm NN} $ &
$\sqrt{(1 + f_1^2)(1 - f_2^2)}$ & $\frac{1 - f_1^2}{1 + f_1^2}$&$\frac{1}{\sqrt{1 + f_1^2}}$ \\
\hline

$(a, b)$ & $(0, c)$ & $a, b, c \in {\rm NN},$ & $\sqrt{(1 - f_1^2)(1
+ f_2^2)}$ & $\frac{1 + f_1^2 - f_2^2 + f_1^2 f_2^2}{(1 - f_1^2)(1 +
f_2^2)}$
&$\frac{1}{C_F}$ \\
 & & $c = a \, {\rm or}\,b $& & &\\ \hline

$(a, b)$ & $(0, c)$ & $c = a \in {\rm NN}, \, b  \in {\rm DN}$ &
$\sqrt{(1 - f_1^2)(1 + f_2^2)}$ & $\frac{1 + f_1^2 + f_2^2 - f_1^2
f_2^2}{(1 - f_1^2)(1 + f_2^2)}$
&$\frac{[1 + f_2^2 (1 - f_1^2)]^{\frac{1}{2}}}{C_F}$ \\
\hline

$(a, b)$ & $(0, c)$ & $c,  a \in {\rm NN}, b \in {\rm DN},$ &
$\sqrt{(1 - f_1^2)(1 + f_2^2)}$ & $\frac{1 + f_1^2}{1 - f_1^2}$
&$\frac{1}{\sqrt{1 - f_1^2}}$ \\
& & $c \neq a $ & & &\\ \hline

 $(a, b)$ & $(c, d)$ & $ a, b, c, d \in {\rm NN},$  &
$\sqrt{(1 + f_1^2)(1 + f_2^2)}$ & $\frac{(1 - f_1^2)(1 - f_2^2) + 4
f_1 f_2}{(1 + f_1^2)(1 + f_2^2)}$&
$\frac{|1 + f_1 f_2|}{C_F}$ \\
  &  & $a = c, b = d $& & &\\
\hline

$(a, b)$ & $(c, d)$ & $ a, b, c, d \in {\rm NN},$
 &
$\sqrt{1 + f_1^2)(1 + f_2^2)}$ & $\frac{1 - f_1^2 - f_2^2 - f_1^2 f_2^2}{(1 + f_1^2)(1 + f_2^2)}$&$\frac{1}{C_F}$ \\
  &  & $a = c, b \neq d,\,{\rm or}\, a = d$ &  & & \\
\hline

$(a, b)$ & $(c, d)$ & $ a, c, d \in {\rm NN},\, b \in {\rm DN},$   &
$\sqrt{1 + f_1^2)(1 + f_2^2)}$ & $\frac{1 - f_1^2 + f_2^2 + f_1^2
f_2^2}{(1 + f_1^2)(1 + f_2^2)}$
&$\frac{[1 + f_2^2(1 + f_1^2)]^{\frac{1}{2}}}{CF}$ \\
 &  & $a =c \, {\rm or}\, d$& & &\\ \hline

$(a, b)$ & $(c, d)$ & $ a, c, d \in {\rm NN}, b \in {\rm DN},$   &
$\sqrt{1 + f_1^2)(1 + f_2^2)}$ & $\frac{1 - f_1^2 }{1 + f_1^2}$ &$\frac{1}{\sqrt{1 + f_1^2}}$ \\
 &  & $a
\neq c, d$& & &\\ \hline
\end{tabular}
\end{table}

\begin{table}
\caption{The cases in Class II} \vskip 0.2cm
\begin{tabular}{|c|c|c|c|c|c|c|c|}
\hline

$(\alpha, \beta)$& $(\gamma, \delta)$ & Index relation & $C_F$ &
$\frac{\lambda_1 + \lambda_1^{-1}}{2}$& $\frac{\lambda_2 + \lambda_2^{-1}}{2}$& $\cos\pi \nu_1$& $\cos\pi\nu_2$\\
\hline

$(a, b)$ & $(0, c)$ & $a, b, c \in {\rm NN},$  & $\sqrt{(1 -
f_1^2)(1 + f_2^2)}$ & $\frac{1 + f_1^2}{1 - f_1^2}$& $\frac{1 -
f_2^2}{1 + f_2^2}$&
$\frac{1}{(1 - f_1^2)^{\frac{1}{2}}}$& $\frac{1}{(1 + f_2^2)^{\frac{1}{2}}}$\\
&  & $ c \neq a, b $ & & & & &\\ \hline

$(a, 0)$ & $(c, d)$ & $a, c, d \in {\rm NN},$  & $\sqrt{(1 +
f_1^2)(1 - f_2^2)}$ & $\frac{1 - f_1^2}{1 + f_1^2}$& $\frac{1 +
f_2^2}{1 - f_2^2}$&
$\frac{1}{(1 + f_1^2)^{\frac{1}{2}}}$& $\frac{1}{(1 - f_2^2)^{\frac{1}{2}}}$\\
&  & $ a \neq c, d $ & & & & &\\ \hline

$(a, b)$ & $(c, d)$ & $a, b, c, d \in {\rm NN},$  & $\sqrt{(1 +
f_1^2)(1 + f_2^2)}$ & $\frac{1 - f_1^2}{1 + f_1^2}$& $\frac{1 -
f_2^2}{1 + f_2^2}$& $\frac{1}{(1 + f_1^2)^{\frac{1}{2}}}$&$\frac{1}{(1 + f_2^2)^{\frac{1}{2}}}$ \\
& & $a, b \neq c, d$ & & & & &\\\hline

\end{tabular}
\end{table}

The Class III matrix elements for matter fields are
\begin{equation}
\label{5eq11} A^X = C_F V_{p-1} e^{- \frac{Y^2}{2\pi \alpha' t}}
\left(2\pi^2 \alpha'\, t\right)^{- \frac{9 - p}{2}} \, \prod_{n =
1}^\infty \frac{1}{(1 - |z|^{2n})^6}\prod_{j = 1}^2 \frac{1}{(1 -
\lambda_j |z|^{2n})(1 - \lambda_j^{-1} |z|^{2n})}
\end{equation}
for both NS-NS and R-R sectors,
\begin{equation}
\label{5eq12} A^\psi_{NS}(\eta',\eta) =\prod_{n = 1}^\infty (1 +
\eta' \eta |z|^{2n - 1})^6 \prod_{j = 1}^2(1 + \eta' \eta\lambda_j
|z|^{2n - 1}) (1 + \eta' \eta \lambda_j^{-1}\, |z|^{2n - 1})
\end{equation}
for the NS-NS sector, and
\begin{eqnarray}
\label{5eq13} A^\psi_{R}(\eta',\eta)A^{\beta\gamma}_{R}(\eta',\eta)
=-(\sqrt{2})^8|z|^2D_F\delta_{\eta' \eta,+}\prod_{n = 1}^\infty (1
+|z|^{2n})^4 \prod_{j = 1}^2(1 + \lambda_j |z|^{2n}) (1 +
\lambda_j^{-1}\, |z|^{2n})
\end{eqnarray}
for the R-R sector. Note that now the R-R sector has non-zero
contribution and we don't separate the contributions from the matter
field $\psi^\mu$ and the superghosts as mentioned earlier. The total
amplitude is now
\begin{eqnarray} \label{7eq3}
\Gamma_{\rm III} = \Gamma_{\rm NS} + \Gamma_{\rm R} &=&\frac{2n_{p
-2}\,n_p \, V_{p-1}\tan\pi\nu_1\,
}{(8\pi^2\alpha^{\prime})^{\frac{p-1}{2}}} \int_{0}^{\infty}dt
e^{-\frac{y^2}{2\pi\alpha^{\prime}t}}t^{-\frac{9-p}{2}}\frac{\theta^2_1(\frac{\nu_1-\nu_2}{2}|it)
\theta^2_1(\frac{\nu_1+\nu_2}{2}|it)}{\eta^6(it)\theta_1(\nu_1|it)\theta_1(\nu_2|it)},
\end{eqnarray}
where again $|z| = e^{- \pi t}$ and $\lambda_j = e^{2 i \pi \nu_j}$
with $j = 1, 2$. Once again in obtaining the above compact
expression, we have expressed the integrand in terms of various
$\theta$-functions and the Dedekind $\eta$-function, then used yet
another fundamental Jacobian identity $2\, \theta^2_1 (\frac{\nu_1 -
\nu_2}{2} | it)$ $ \theta_1^2 (\frac{\nu_1 + \nu_2}{2} | it) =
\theta_3 (\nu_1 | it)\, \theta_3 (\nu_2 | it)\, \theta_3^2 (0 | it)
- \theta_4 (\nu_1 | it)\, \theta_4 (\nu_2 | it) \, \theta^2_4 (0 |
it) - \theta_2 (\nu_1 | it)\, \theta_2 (\nu_2 | it)\, \theta_2^2 (0
| it)$ where the first two terms come from the NS-NS sector and the
last term comes from the R-R sector. This identity is also a special
form of (iv) given on page 468 in \cite{whittaker-watson}. The
constants $C_F$ and the sum of $\lambda_j + \lambda_j^{-1}$ for
cases considered in this class are listed in Table 3 with $D_F = -
f_2/C_F = \cos\pi\nu_1 \cos\pi\nu_2$.

\begin{table}
\caption{The cases in Class III} \vskip 0.2cm
\begin{tabular}{|c|c|c|c|c|c|c|c|}
\hline

$(\alpha, \beta)$& $(\gamma, \delta)$ & Index Relation & $C_F$ &
$\frac{\lambda_1 + \lambda_1^{-1}}{2}$& $\frac{\lambda_2 + \lambda_2^{-1}}{2}$&$\cos\pi\nu_1$&$\cos\pi\nu_2$ \\
\hline

$(a, b)$ & $(0, c)$ & $ c \in {\rm NN},$   & $\sqrt{(1 - f_1^2)(1 +
f_2^2)}$ & $\frac{1 + f_1^2}{1 - f_1^2}$& $ - \frac{1 - f_2^2}{1 +
f_2^2}$
& $\frac{1}{(1 - f_1^2)^{\frac{1}{2}}}$& $- \frac{f_2}{(1 + f_2^2)^{\frac{1}{2}}}$\\
& & $a, b \in {\rm DN}$ & & & & & \\ \hline

$(a, b)$ & $(c, d)$ & $ c, d \in {\rm NN},$  & $\sqrt{(1 + f_1^2)(1
+ f_2^2)}$ & $\frac{1 - f_1^2}{1 +
f_1^2}$ & $ - \frac{1 - f_2^2}{1 + f_2^2}$& $\frac{1}{(1 + f_1^2)^{\frac{1}{2}}}$& $- \frac{f_2}{(1 + f_2^2)^{\frac{1}{2}}}$ \\
& & $a, b \in {\rm DN}$ & & & & & \\ \hline

\end{tabular}
\end{table}
We now come to discuss the nature and range of $\nu'$s for cases in
each class which will depend crucially on the nature of fluxes
(electric or magnetic) involved as discussed in
\cite{Lu:2009yx,Lu:2009kv2}. Note that for an electric flux $f$,  $
0 < |f| < 1$ with the critical field $|f| = 1$ while for a magnetic
flux $f$,  $0 < |f| < \infty$. In Class I, when both fluxes are
electric, $\nu$ is imaginary, i.e., $\nu = i \nu_0$ and $\cos\pi\nu
= \cosh\pi\nu_0$ with $0 < \nu_0 < \infty$ (see footnote 6 for
detail). When the two are both magnetic, $\nu = \nu_0$ is real with
$ 0 < \nu_0 < 1/2$. When one flux is electric and the other
magnetic, we have the following cases: 1) $(\hat F)_{0a}$ and $(\hat
F')_{cd}$ with $a \in {\rm DN}, \, c, d \in {\rm NN}$, then $\nu =
\nu_0$ is real with $0 < \nu_0 < 1/2$; 2) $(\hat F)_{ab}$ and $(\hat
F')_{0c}$ with $c, a \in {\rm NN},\, b \in {\rm ND}$, then $\nu = i
\nu_0$ (also $\cos\pi\nu = \cosh\pi\nu_0$) is imaginary with $0 <
\nu_0 < \infty$; 3) $(\hat F)_{0a}$ and $(\hat F')_{cd}$ with $a, c,
d \in {\rm NN},\, a = c \, {\rm or}\, d$ (or $(\hat F)_{ab}$ and
$(\hat F')_{0c}$ with $a, b, c \in {\rm NN},\, c = a \, {\rm or}\,
b$), $\cos\pi \nu = 1/\sqrt{(1 + f_1^2)(1 - f_2^2)}$ (or $ =
1/\sqrt{(1 - f_1^2)(1 + f_2^2)}$) where $\nu$ can be non-vanishing
real, imaginary, or zero, depending on $\sqrt{(1 - f_1^2)(1 +
f_2^2)} (\,{\rm or}\,\sqrt{(1 + f_1^2)(1 - f_2^2)}\,) > 1, < 1
\,{\rm or}\, = 1$, respectively, for non-vanishing
fluxes\footnote{Here when $\nu$ is real, we call it magnetically
dominant otherwise electrically dominant.}. For Class II, the nature
of $\nu_j$ ($j = 1, 2$) is directly related to the corresponding
flux and  $\nu_j = i \nu_{j0}$ is imaginary with $0 < \nu_{j0} <
\infty$ if the corresponding flux is electric and $\nu_j = \nu_{j0}$
is real with $0 < \nu_{j0} < 1/2$ if the flux is magnetic. For Class
III, while $\nu_1$ remains the same nature as the $\nu_j$ in Class
II, $\nu_2 = \nu_{20}$ is however always real with now $0 < \nu_{20}
< 1$ for which $0 < \nu_{20} < 1/2$ corresponds to $f_2 < 0$,
$\nu_{20}= 1/2$ to $f_2 = 0$ and $1/2 < \nu_{20} < 1$ to $f_2 > 0$.

Let us first consider the large-separation limit of the above
amplitudes for $2 \le p \le 6$. This amounts to taking  the large
t-limit for the $\theta_1$-function and the Dedekind $\eta$-function
in each integrand, i.e., $ \theta_1(\nu|it)\rightarrow
2\,e^{-\frac{\pi t}{4}}\sin\pi\nu,\,\eta(it)\rightarrow
e^{-\frac{\pi}{12}t}$ (noting now $|z|=e^{-\pi t}\rightarrow0$). We
then have $\label{4eq28} \Gamma_{\rm I} \rightarrow C_{\rm I} /Y^{7
- p}$ where a simple integration has been performed and  $C_{\rm I}
= n_p\,n_{p - 2} \, c_p \,c_{p - 2} \,V_{p - 1} C_F (\lambda +
\lambda^{-1} + 2) /[4(7 - p) \Omega_{8 - p}]$ with
$(7-p)\Omega_{8-p} = 4\pi\pi^{(7-p)/2} /$ $ \Gamma((7-p)/2)$ and
$\Omega_q$ the volume of unit $q$-sphere. Similarly, we have $
\Gamma_{\rm II} \rightarrow C_{\rm II}/Y^{7 - p}$ with $C_{\rm II} =
n_p n_{p - 2}\,  c_p c_{p - 2}\, V_{p-1} C_F
(\lambda_1+\lambda^{-1}_1+\lambda_2+\lambda^{-1}_2)/[4 (7 - p)
\Omega_{8-p}]$, and $\Gamma_{\rm III} \rightarrow C_{\rm III}/Y^{7 -
p}$ with $C_{\rm III} = n_p n_{p -2} \, c_p c_{p - 2} \, V_{p-1}
C_F\,(\lambda_1+\lambda^{-1}_1+\lambda_2+\lambda^{-1}_2+4-8D_F) /[4
(7-p)\Omega_{8-p}]$. Note that the parameters $\lambda,\,{\rm
or}\,\, \lambda_j (j = 1, 2),\, C_F ({\rm and}\,D_F)$ in each
relevant class are given in the respective table and $c_k =
\sqrt{\pi} (2\pi \sqrt{\alpha'})^{3 - k}$ with $k = p, p-2$.

Before proceeding, we pause to discuss which amplitude can vanish
when either flux is non-critical and non-vanishing. Note that the
vanishing large-separation amplitude also implies the vanishing of
the corresponding amplitude at a general separation. This can
possibly occur only for one case in each class and only when both
fluxes are magnetic. Concretely, the amplitude vanishes in Class I
for the case of $a, b, c, d \in {\rm NN}, a, b = c, d$ (see Table 1)
when $f_1 f_2 = -1$ (corresponding to $\nu = 1/2$),  in Class II for
the case of $a, b, c, d \in {\rm NN}, a, b \neq c, d$ (see Table 2)
when $f_1 f_2 = \pm 1$ (corresponding to $\nu_1 + \nu_2 = 1/2$), and
in Class III for the case of  $a, b \in {\rm DN}, c, d \in {\rm NN}$
(see Table 3) when $f_1 f_2 = \pm 1$ and $ f_2 < 0$ (corresponding
to $\nu_1 = \nu_2$). The vanishing of amplitude also implies the
preservation of certain number of supersymmetries for the underlying
system which can be similarly analyzed following the steps given in
the appendix of \cite{Lu:2009kv2}. Here we simply state the results:
the case in Class I or III preserves 1/4 of the spacetime
supersymmetries while in Class II it preserves only 1/8 of the
supersymmetries.

Apart from the above three special cases and one case associated
with the special case in Class II, all the other large-separation
amplitudes are positive, therefore giving attractive interactions.
This particular case corresponds to the one given in Table 2 in
Class II when both fluxes are magnetic with their respective spatial
indices $a, b, c, d \in {\rm NN};\, a, b \neq c, d$ and $f_1^2 f_2^2
> 1$ (corresponding to $1/2 < \nu_1 + \nu_2 < 1$), and gives a negative
amplitude, hence a repulsive interaction. For the magnetic or
magnetically dominating (in the sense mentioned earlier) cases, the
nature of interaction will keep hold even at a general separation.
However, the nature of force will become indefinite  at small
separation  when the effects of fluxes are electric or electrically
dominant. The basic feature about the nature of interaction on the
brane-separation remains similar to our earlier discussion for
systems with $p = p'$ considered in \cite{Lu:2009yx,Lu:2009kv2}.

We now move to discuss the analytical structure of amplitudes at
small separation $Y$ for which the open string description is
appropriate. This can be achieved via the Jacobian transformation of
the integration variable $t \rightarrow t' = 1/t$, converting the
tree-level closed string cylinder diagram to the open string
one-loop annulus diagram.  In terms of this annulus variable
$t^{\prime}$, noting \bea \eta (\tau) = \frac{1}{\left(- i
\tau\right)^{1/2}} \eta \left(- \frac{1}{\tau}\right),\qquad
\theta_1 (\nu|\tau) = i \frac{\,e^{- i \pi \nu^2/\tau}}{\left(- i
\tau\right)^{1/2}} \theta_1 \left(\left.\frac{\nu}{\tau}\right|-
\frac{1}{\tau}\right),\eea the amplitude in each class can be
explicitly re-expressed, respectively, as
\begin{eqnarray} \label{4eq32}
\Gamma_{\rm I} &=&-\frac{2\, n_p n_{p - 2} \,
V_{p-1}C_F\sin\pi\nu}{(8\pi^2\alpha^{\prime})^{\frac{p-1}{2}}}
\int_{0}^{\infty}dt^{\prime} e^{-\frac{Y^2
t^{\prime}}{2\pi\alpha^{\prime}}}t^{\prime\frac{1-p}{2}}
\frac{\left[\cos (-i\pi \nu t') - \cosh\frac{\pi
t'}{2}\right]^2}{\sin(-i\pi\nu t^{\prime})\sin(-i\pi
t^{\prime}/2)}\nn &
&\times\prod_{n=1}^{\infty}\frac{1}{\left(1-|z|^{2n}\right)^4}
\prod_{j=1}^{2}\frac{\left(1-e^{\pi(\nu+(-)^j/2)t^{\prime}}
|z|^{2n}\right)^2\left(1-e^{-\pi(\nu+(-)^j/2)t^{\prime}}|z|^{2n}\right)^2}
{\left(1-e^{(-)^{j}2\pi\nu
t^{\prime}}|z|^{2n}\right)\left(1-e^{(-)^{j}\pi
t^{\prime}}|z|^{2n}\right)},
\end{eqnarray}

\begin{eqnarray} \label{4eq33}
\Gamma_{\rm II}  &=&-\frac{2\,n_p n_{p - 2} \,
V_{p-1}\tan\pi\nu_1\tan\pi\nu_2}{(8\pi^2\alpha^{\prime})^{\frac{p-1}{2}}}
\int_{0}^{\infty}dt^{\prime} e^{-\frac{Y^2
t^{\prime}}{2\pi\alpha^{\prime}}}t^{\prime\frac{3-p}{2}}\nn &
&\times\frac{\left[\cos \left(- i\pi (\nu_1 - \nu_2) t'\right) -
\cosh \frac{\pi t'}{2}\right]\left[\cos\left(- i \pi (\nu_1 + \nu_2)
t'\right) - \cosh \frac{\pi t'}{2}\right] }{\sin(-i\pi\nu_1
t^{\prime})\sin(-i\pi\nu_2 t^{\prime})\sinh(\pi
t^{\prime}/2)}\prod_{n=1}^{\infty}
\frac{1}{\left(1-|z|^{2n}\right)^2}\nn & &\times\prod_{k = 1}^2
\frac{1}{1-e^{(-)^k\pi t^{\prime}}|z|^{2n}}\prod_{j = 1}^{2}\frac{1-
2 |z|^{2n} \cosh [\pi(\nu_1+(-)^j\nu_2+ \frac{(-)^k}{2})t^{\prime}]
+ |z|^{4n}} {1-e^{(-)^k2\pi\nu_j t^{\prime}}|z|^{2n}},
\end{eqnarray}

\begin{eqnarray} \label{4eq34}
\Gamma_{\rm III} &=&-\frac{2 n_p n_{p - 2}
V_{p-1}\tan\pi\nu_1}{(8\pi^2\alpha^{\prime})^{\frac{p-1}{2}}}
\int_{0}^{\infty}dt^{\prime} e^{-\frac{Y^2
t^{\prime}}{2\pi\alpha^{\prime}}}t^{\prime\frac{1-p}{2}}\frac{\left[\cos(-
i \pi \nu_1 t' ) - \cos(- i \pi \nu_2 t') \right]^2}{\sin(-i\pi\nu_1
t^{\prime})\sin(-i\pi\nu_2 t^{\prime})}\nn &
&\times\prod_{n=1}^{\infty}\frac{1}{\left(1-|z|^{2n}\right)^4}
\prod_{j=1}^{2}\frac{\left(1-e^{\pi(\nu_1+(-)^j\nu_2)t^{\prime}}
|z|^{2n}\right)^2\left(1-e^{-\pi(\nu_1+(-)^j\nu_2)t^{\prime}}|z|^{2n}\right)^2}
{\left(1-e^{2\pi\nu_jt^{\prime}}|z|^{2n}\right)\left(1-e^{-2\pi\nu_jt^{\prime}}|z|^{2n}\right)},
\end{eqnarray}
with now $|z|=e^{-\pi t^{\prime}}$. When $\nu = \nu_0$ in Class I
with $0 < \nu_0 < 1/2$ or  $\nu_j = \nu_{j0}$  with $0 < \nu_{j0} <
1/2$ ($j = 1, 2$) in Class II or with $0 < \nu_{10} < 1/2$ and $0 <
\nu_{20} < 1$ in Class III, the underlying amplitude remains real
and will diverge when the separation $Y \leq \pi\sqrt{2 ( 1/2 -
\nu)\alpha'}$ in Class I, or $Y \leq \pi\sqrt{2 (1/2 - \nu_1 -
\nu_2)\alpha'}$ when $0 < \nu_1 + \nu_2 < 1/2$ in Class II, or $Y
\leq \pi\sqrt{2|\nu_1 - \nu_2|\alpha'} $ in Class III when $\nu_1
\neq \nu_2$, i.e., with each on the order of string scale. Along
with a similar discussion given in
\cite{Lu:2009yx,Lu:2009kv2,Banks:1995ch,Lu:2007kv}, this divergence
indicates the onset of tachyonic instability in each respective
case, giving rise to the relaxation of the underlying system to form
the final stable bound state. When $1/2 < \nu_1 + \nu_2 < 1$ in
Class II, the force as mentioned earlier is repulsive and becomes
larger when the separation $Y$ becomes smaller and for this reason
the integrand has no exponential blow-up singularity to show up even
at $t' \rightarrow \infty$, i.e., no onset of tachyonic singularity,
for any $Y > 0$. Note that if we express the amplitude \eqn{4eq34}
in Class III above in terms of $p' = p - 2$, i.e., the spatial
NN-directions, its structure looks similar to the one given in
\cite{Lu:2009kv2} for $p = p'$ with the respective two fluxes not
sharing any common index. So many of the underlying properties such
as the onset of various instabilities remain the same as those given
in \cite{Lu:2009kv2}, therefore referred there for detail. For this
reason, we discuss the remaining new features only below for this
Class.

We move to discuss the more rich structure and the associated
physics when  $\nu = i\nu_0$ with $0 < \nu_0 < \infty$ in Class I or
$\nu_1 = i \nu_{10}$ with $0 < \nu_{10} < \infty$ and $\nu_{2} =
\nu_{20}$ with $\nu_{20}$ real in Class II\footnote{The case with
$\nu_1$ real and $\nu_2$ imaginary can be similarly discussed in
this class.} or III .  In Class I, this corresponds to the presence
of at least one electric flux (or being electrically dominant) along
a NN-direction. Now the integrand has an infinite number of simple
poles occurring on the positive real axis at $t' = k/\nu_0$ with $k
= 1, 2, \cdots$ and for this, as discussed in
\cite{Bachas:1992bh,Bachas:1992zr,Green:1996um,Cho:2005aj,Porrati:1993qd,
Chen:2008kv,Lu:2009yx,Lu:2009kv2}, the amplitude has an imaginary
part which is sum of the residues at these simple poles and gives
rise to the rate of pair production of open strings. This rate per
unit $(p - 1)$-worldvolume is
\begin{eqnarray} \label{4eq38}
{\cal W}_{\rm I} \equiv-\frac{2 {\rm Im} \Gamma_{\rm I}}{V_{p-1}}
&=& \frac{4 n_p n_{p - 2} C_F
\sinh\pi\nu_0}{\nu_0(8\pi^2\alpha^{\prime})^{\frac{p-1}{2}}}\sum_{k
= 1}^\infty (-)^{k+1}\left(\frac{\nu_0}{k}\right)^{\frac{p-1}{2}}
e^{- \frac{k Y^2 }{2 \pi \nu_0 \alpha' }} \frac{\left[\cosh
\frac{k\pi}{2\nu_0}- (-)^k\right]^2}{\sinh\frac{k\pi}{2\nu_0}}\nn &
&\times\prod_{n = 1}^\infty \frac{\left[1 - 2 (-)^k  e^{- \frac{2n k
\pi}{\nu_0}}\cosh\frac{\pi k}{2\nu_0} + e^{- \frac{4 n k
\pi}{\nu_0}}\right]^4} { \left(1 - e^{- \frac{2 n k\pi}{\nu_0
}}\right)^6\left(1 - 2 e^{- \frac{2n k\pi}{\nu_0}} \cosh\frac{\pi
k}{\nu_0} + e^{- \frac{4 n k \pi}{\nu_0}} \right)}.
\end{eqnarray}
This rate shares the common features as found in similar cases
studied in \cite{Lu:2009kv2} when $p = p'$ with the two fluxes
sharing at least one common index. Namely, it is suppressed by the
separation $Y$ but enhanced by the value of $\nu_0$, which can be
determined by flux/fluxes via $\cosh\pi\nu_0 = \cos\pi\nu$ as given
Table 1. Each term in the sum is suppressed by the integer $k$ but
 diverges
as the electric field reaches its critical value, i.e., $\nu_0
\rightarrow \infty$, signaling the onset of a singularity
\cite{Porrati:1993qd}. However, the present rate differs in many
aspects: the number of terms in the sum doubles and the contribution
to the rate is positive for odd $k$ and negative for even $k$. In
particular, there is an enhanced factor $[\cosh \frac{k\pi}{2\nu_0}-
(-)^k ]^2/\sinh\frac{k\pi}{2\nu_0}$ which becomes important for
small $\nu_0$. This can be examined easily by looking at the leading
order
 approximation, i.e., the $k = 1$ term in the sum as
\begin{eqnarray} \label{4eq39} (2\pi\alpha')^{(p' + 1)/2} {\cal W}_{\rm I} \approx 2\pi n_p
n_{p - 2} C_F \left(\frac{\nu_0}{4\pi}\right)^{(p-1)/2}
e^{-\frac{Y^2-\pi^2\alpha^\prime}{2\pi \nu_0\alpha^\prime}},
\end{eqnarray} where $C_F$ is given in Table 1 and the small $\nu_0$
can be explicitly determined, to leading order, in terms of the
fluxes present. Note that small $\nu_0$ does need all fluxes along
NN directions, electric or magnetic, to be small. Let us consider
the first case in Table 1 as an illustration for which $C_F \approx
1$. Now $\cosh\nu_0 = \cos\nu = (1 - f_1 f_2)/\sqrt{(1 - f_1^2)(1 -
f_2^2)}$ which gives, to leading order, $\nu_0 \approx |f_1 -
f_2|/\pi$. Since $p \ge 3$, the rate is largest for $p = 3$ for
given $C_F$ and small $\nu_0$ (for fixed $n_p$ and $n_{p - 2}$) and
can be significant at a separation $Y = \pi \sqrt{\alpha'} + 0^+ $,
i.e. on the order of string scale, where we don't have yet the onset
of tachyonic instability which occurs for $Y \le \pi\sqrt{\alpha'}$
from the real part of the amplitude. This is in spirit similar to
the enhanced rate discussed in \cite{Lu:2009kv2}, for which $p = p'$
with $p = p' = 3$ giving the largest rate, but for a completely
different case where a reasonably large magnetic flux not sharing
any common index with the weak electric flux must be present.  The
novel feature here is that the $p' \,( = p - 2$) branes play
effectively as such a magnetic flux and this is rational for the
enhancement even in the absence of a magnetic flux. As discussed
above, for small $\nu_0$, only the magnetic flux (electric flux)
with one index along a DN direction can further enhance (reduce) the
rate through $C_F \approx \sqrt{1 + f^2}\, ( \approx \sqrt{1 - f^2})
$ with $f$ the magnetic (electric) flux. Note that the magnetic flux
$|f|$ is measured in string units and for a realistic value, $|f|$
should be smaller than unity, therefore giving $C_F \approx 1$. In
other words, the enhancement due to a magnetic flux in general is
small and so we can ignore this for Class I from now on.  Let us
make some numerical estimation of this rate for small $\nu_0$ and
this may serve for sensing its significance. For this purpose, we
take $n_p = n_{p - 2} = 5$, $\nu_0 = 0.02$ and $C_F \approx 1$. We
also take the brane separation as given above such that the
exponential in the rate can be approximated to one. So the rate in
string units is $(2\pi \alpha')^{(p' + 1)/2} {\cal W}_I = 2\pi n_{p
- 2} n_p (\nu_0/4\pi)^{(p - 1)/2} = 0.25, 0.02$ for $p = 3, 4$,
respectively. The largest rate occurs indeed at $p = 3$, the rate
for $p = 4$ is one order of magnitude smaller and the larger the $p$
the smaller the rate. This estimation indicates that the rate for $p
= 3$ can indeed be significant for small separation,  for reasonably
chosen $n_p$ and $n_{p - 2}$, and even for small fluxes, therefore
more realistic than the case mentioned above in \cite{Lu:2009kv2}.
The small-separation also implies that the significantly produced
open string pairs are almost confined on the branes along the
electric flux line. This further implies that the radiations due to
the annihilation of the open string pairs in a short time should be
mostly along the brane directions. If string theories are relevant,
given the large rate for $p = 3$, we expect that the early Universe
or even macroscopic objects in the sky at present can give rise to
such open string pair production, therefore large radiations, which
may have potential observational consequence. This may further imply
why our world has three large spatial dimensions since the
observational consequence if any can only be significant for $p =
3$. Pursuing any of these possibilities is beyond the scope of
present work and we expect to examine carefully some of these
possibilities in the future.

We have also learned from the above that we need at least one
electric flux being along the NN-direction to give rise to the pair
production. This is clearly indicated in the case of $(\alpha,
\beta) = (0, a), (\gamma, \delta) = (c, d), a \in {\rm DN}, c, d \in
{\rm NN}$ given in Table 1 for which the electric flux is along a
spatial DN-direction and as such $\nu = \nu_0$ is real (therefore no
pair production of open strings), depending only on the magnetic
flux. The triviality of the electric flux in this case can be
understood via a T-duality along the electric flux direction. The
D$_p$ branes then become D$_{p - 1}$ branes while the D$_{p'}$
become D$_{p' + 1}$ with the D$_{p - 1}$ moving with a velocity,
determined by the original electric flux along the T-dual direction,
relative to the D$_{p' + 1}$ in the T-dual picture. Since the D$_{p'
+ 1}$ has a Lorentz symmetry along the T-dual direction, so such a
relative motion can be  removed by a Lorentz boost along this
direction. If we now T-dual back, we end up with a system without
the presence of an electric flux but with an additional overall
factor in the amplitude due to the boost.

 For Class II or III, there is only
one case relevant (see footnote 8 for Class II) for which $\nu_1 = i
\nu_{10}$ is imaginary ($0 < \nu_{10} < \infty$) while $\nu_2 =
\nu_{20}$ is real with $ 0 < \nu_{20} < 1/2$ in Class II and with
$0< \nu_{20} < 1$ in Class III. This corresponds to $(\alpha, \beta)
= (a, b), (\gamma, \delta) = (0, c)$ with $c \in {\rm NN}$, i.e.,
the electric flux along a NN-direction and in Class II, $c \neq a, b
\in {\rm NN}$ (see Table 2) while $a, b \in {\rm DN}$ in Class III
(see Table 3). In either case, the integrand has also an infinite
number of simple poles occurring on the positive real axis at $t' =
k/\nu_{10}$ with $k = 1, 2, \cdots$. By the same token as in Class I
above, the pair production rate for the case in Class II ($p \ge 5$)
is
\begin{eqnarray}
\label{rate2} &&{\cal W}_{\rm II} = \frac{4 n_p n_{p - 2}
\tanh\pi\nu_{10} \tan \pi\nu_{20}}{\nu_{10} (8 \pi^2
\alpha')^{\frac{p - 1}{2}}} \sum_{k = 1}^\infty (-)^{k + 1}
\left(\frac{\nu_{10}}{k}\right)^{\frac{p - 3}{2}}  \frac{\left[(-)^k
\cosh\frac{\pi\nu_{20} k}{\nu_{10}} - \cosh\frac{\pi k}{2
\nu_{10}}\right]^2} {\sinh\frac{\pi\nu_{20} k}{\nu_{10}}
\sinh\frac{\pi k}{2 \nu_{10}}}\nn && \times e^{- \frac{Y^2
k}{2\pi\nu_{10} \alpha'}} \prod_{n = 1}^\infty \frac{1}{(1 -
e^{-\frac{2 n k\pi}{\nu_{10}}})^4}\prod_{j = 1}^2 \frac{\left[1 - 2
(-)^k e^{- \frac{2nk\pi}{\nu_{10}}} \cosh\frac{\pi
k}{\nu_{10}}\left(\nu_{20} + \frac{(-)^j}{2}\right) + e^{-
\frac{4nk\pi}{\nu_{10}}}\right]^2}{\left(1 - e^{-
\frac{k\pi}{\nu_{10}}(2n + (-)^j)}\right)\left(1 - e^{-
\frac{2nk\pi}{\nu_{10}}(n + (-)^j \nu_{20})}\right)}.
\end{eqnarray}
This case looks in almost every aspect, for examples, the onset of
various instabilities for both real and imaginary parts of the
amplitude, even closer to the one in \cite{Lu:2009kv2} with the two
fluxes not sharing any common index, and for this reason not
repeating here, except for one subtle point regarding the
enhancement factor for the pair production rate given above. Let us
specify this for small $\nu_{10}$ (for fixed non-vanishing
$\nu_{20}$) for which the rate can be approximated by the leading $k
= 1$ term as
\begin{equation} (2\pi\alpha')^{(p' + 1)/2} {\cal W}_{\rm II} \approx  n_p n_{p -
2} \left(\frac{\nu_{10}}{4 \pi}\right)^{(p' - 1)/2} e^{- \frac{Y^2
}{2\pi\nu_{10} \alpha'}} e^{\frac{\pi(1 - 2 \nu_{20})}{2
\nu_{10}}}\tan\pi \nu_{20}.\end{equation}  If we focus on the
NN-directions, i.e., $p' = p - 2 \ge 3$, the dimensionless rate
$(2\pi\alpha')^{(p' + 1)/2} {\cal W}_{\rm II}$ differs from the
first equality given in Eq.(87) in \cite{Lu:2009kv2} only in the
enhancement factor and the ratio of the two is $e^{\frac{\pi(1 - 4
\nu_{20})}{2 \nu_{10}}} > 1$
 for $0 < \nu_{20} < 1/4$ and is less  than unity for $1/4 <
\nu_{20} < 1/2$. For $\nu_{20} \ll \nu_{10}$, i.e., vanishing
magnetic flux, the rate looks identical to the one given in
Eq.(\ref{4eq39}) above in Class I but with now $p \ge 5$ and $C_F$
set to unity, with again the D$_{p'}$ as an effective magnetic flux
mentioned earlier.

For the case in Class III, \bea \label{rate3} {\cal W}_{\rm III} &=&
\frac{4 n_p
 n_{p - 2} \tanh \pi\nu_{10}}{\nu_{10}} \sum_{k =
1}^\infty (- )^{k + 1} \left(\frac{\nu_{10}}{8 k \pi^2
\alpha'}\right)^{\frac{p - 1}{2}} e^{- \frac{k Y^2}{2\pi \nu_{10}
\alpha'}} \frac{\left[\cosh \frac{k \pi \nu_{20}}{\nu_{10}} -
(-)^k\right]^2}{ \sinh \frac{k\pi \nu_{20}}{\nu_{10}}} \nn
&\,&\qquad\times \prod_{n = 1}^\infty \frac{\left[1 - 2 (-)^k e^{-
\frac{2 n k \pi}{\nu_{10}}} \cosh \frac{k\pi \nu_{20}}{\nu_{10}} +
e^{- \frac{4 n k \pi}{\nu_{10}}}\right]^4}{\left[1 - e^{- \frac{2 n
k \pi}{\nu_{10}}}\right]^6 \left[1 - e^{- \frac{2 k \pi}{\nu_{10}}(n
- \nu_{20})}\right]\left[1 - e^{- \frac{2 k \pi}{\nu_{10}}(n +
\nu_{20})}\right]},\eea which again shares many common features
mentioned above in Class II as the one given in \cite{Lu:2009kv2}.
For small $\nu_{10}$, the rate can be approximated by the leading $k
= 1$ term as \be (2\pi \alpha')^{(p' + 1)/2} {\cal W}_{\rm III} =
2\pi n_p n_{p - 2} \left(\frac{\nu_{10}}{4\pi}\right)^{(p - 1)/2}
e^{- Y^2/2\pi \nu_{10}\alpha'} e^{\pi \nu_{20}/\nu_{10}},\ee with $p
= p' + 2 \ge 3 $. This dimensionless rate is largest for $p = 3$ (or
$p' = 1$) with an enhancement factor $2\pi e^{\pi\nu_{20}/\nu_{10}}$
vs $e^{\pi\nu_{20}/\nu_{10}} \tan\pi\nu_{20}$ given in
\cite{Lu:2009kv2}. Apart from a difference of a factor of
$\tan\pi\nu_{20}$,  the nature and range of $\nu_{20}$ here are
completely different\footnote{In \cite{Lu:2009kv2}, $\nu_{20}
\rightarrow 0$ when the magnetic flux $f_2 \rightarrow 0$ and
$\nu_{20} \rightarrow 1/2$ when $|f_2| \rightarrow \infty$.}. In the
present case, $0 < \nu_{20} < 1$ with $\nu_{20} = 1/2$ corresponding
to vanishing magnetic flux, $\nu_{20} \rightarrow 0$ when the
magnetic flux $f_2 \rightarrow - \infty$ and $\nu_{20} \rightarrow
1$ when $f_2 \rightarrow \infty$. So once again even without the
presence of a magnetic flux, i.e., $\nu_{20} = 1/2$, we still have
an exponential enhancement of the rate and the rate in each Class
looks identical in the absence of a magnetic flux. The various
implications of the rate in class II or III can be similarly
discussed as in Class I above and will not be repeated, except for
one point to which we turn next. The largest rate in Class II occurs
for $p' = 3$ and $p = 5$. So if there is indeed an observational
consequence, this may indicate the existence of large extra
dimensions since $p = 5$ other than the usual $p = 3$ in this case.

\section*{Acknowledgements}
The authors would like to thank the anonymous referee for comments
and suggestions which help us to improve the manuscript, to Rong-Jun
Wu, Zhao-Long Wang, Bo Ning and Ran Wei for useful discussion. We
acknowledge support by grants from the Chinese Academy of Sciences,
a grant from 973 Program with grant No: 2007CB815401 and grants from
the NSF of China with Grant No:10588503 and 10535060.

\appendix
\newcounter{pla}
\renewcommand{\thesection}{\Alph{pla}}
\renewcommand{\theequation}{\Alph{pla}.\arabic{equation}}
\setcounter{pla}{1} \setcounter{equation}{0}


\vspace{.5cm}

\begin{thebibliography}{99}
%\cite{Di Vecchia:1999rh}
\bibitem{Di Vecchia:1999rh}
  P.~Di Vecchia and A.~Liccardo,
  ``D branes in string theory. I,''
  NATO Adv.\ Study Inst.\ Ser.\ C.\ Math.\ Phys.\ Sci.\  {\bf 556}, 1 (2000)
  [arXiv:hep-th/9912161].
  %%CITATION = NASCD,556,1;%%

%\cite{Lu:2009kv2}
\bibitem{Lu:2009kv2}
  J.~X.~Lu, and S.~S.~Xu,
  ``The open string pair-production rate enhancement,''
  [arXiv:0904.4112 [hep-th]].

%\cite{Polchinski: booktwo}
\bibitem{polbooktwo}
  J.~Polchinski, ``Superstring Theory",
  Vol. 2, Cambridge: Cambridge University Press (1998)
%%

%\cite{Lu:2009yx}
\bibitem{Lu:2009yx}
  J.~X.~Lu, B.~Ning, R.~Wei and S.~S.~Xu,
  ``Interaction between two non-threshold bound states,''
  Phys.\ Rev.\  D {\bf 79}, 126002 (2009)
  [arXiv:0902.1716 [hep-th]].
  %%CITATION = PHRVA,D79,126002;%%

%\cite{DiVecchia:1999uf}
\bibitem{DiVecchia:1999uf}
  P.~Di Vecchia, M.~Frau, A.~Lerda and A.~Liccardo,
  ``(F,Dp) bound states from the boundary state,''
  Nucl.\ Phys.\  B {\bf 565}, 397 (2000)
  [arXiv:hep-th/9906214].
  %%CITATION = NUPHA,B565,397;%%

%\cite{Billo:1998vr}
\bibitem{Billo:1998vr}
  M.~Billo, P.~Di Vecchia, M.~Frau, A.~Lerda, I.~Pesando, R.~Russo and S.~Sciuto,
  ``Microscopic string analysis of the D0-D8 brane system and dual R-R
  states,''
  Nucl.\ Phys.\  B {\bf 526}, 199 (1998)
  [arXiv:hep-th/9802088].
  %%CITATION = NUPHA,B526,199;%%


%\cite{Whittaker-watson: book}
\bibitem{whittaker-watson}
E.~ T. Whittaker and G.~N. Watson, ``A Course of Modern Analysis",
4th Ed (reprinted). Cambridge: Cambridge University Press (1963) %%

%\cite{Banks:1995ch}
\bibitem{Banks:1995ch}
  T.~Banks and L.~Susskind,
  ``Brane - Antibrane Forces,''
  arXiv:hep-th/9511194.
  %%CITATION = HEP-TH/9511194;%%


%\cite{Lu:2007kv}
\bibitem{Lu:2007kv}
  J.~X.~Lu, B.~Ning, S.~Roy and S.~S.~Xu,
  ``On Brane-Antibrane Forces,''
  JHEP {\bf 0708}, 042 (2007)
  [arXiv:0705.3709 [hep-th]].
  %%CITATION = JHEPA,0708,042;%%

%\cite{Bachas:1992bh}
\bibitem{Bachas:1992bh}
  C.~Bachas and M.~Porrati,
  ``Pair creation of open strings in an electric field,''
  Phys.\ Lett.\  B {\bf 296}, 77 (1992)
  [arXiv:hep-th/9209032].
  %%CITATION = PHLTA,B296,77;%%

%\cite{Bachas:1992zr}
\bibitem{Bachas:1992zr}
  C.~Bachas,
  ``Schwinger effect in string theory,''
  arXiv:hep-th/9303063.
  %%CITATION = HEP-TH/9303063;%%

%\cite{Green:1996um}
\bibitem{Green:1996um}
  M.~B.~Green and M.~Gutperle,
  ``Light-cone supersymmetry and D-branes,''
  Nucl.\ Phys.\  B {\bf 476}, 484 (1996)
  [arXiv:hep-th/9604091].
  %%CITATION = NUPHA,B476,484;%%
%\cite{Cho:2005aj}
\bibitem{Cho:2005aj}
  J.~H.~Cho, P.~Oh, C.~Park and J.~Shin,
  ``String pair creations in D-brane systems,''
  JHEP {\bf 0505}, 004 (2005)
  [arXiv:hep-th/0501190].
  %%CITATION = JHEPA,0505,004;%%


%\cite{Porrati:1993qd}
\bibitem{Porrati:1993qd}
  M.~Porrati,
  ``Open strings in constant electric and magnetic fields,''
  arXiv:hep-th/9309114.
  %%CITATION = HEP-TH/9309114;%%

%\cite{Chen:2008kv}
\bibitem{Chen:2008kv}
  B.~Chen, and X.~Liu,
  ``D1-D3 (or $\overline{D}_3$) Systems with Fluxes,''
  JHEP {\bf 0808}, 034 (2008)
  [arXiv:0806.3548 [hep-th]].









\end{thebibliography}
\end{document}